\documentclass[a4paper,fleqn]{cas-dc}
\usepackage[sort&compress, numbers]{natbib}
\usepackage{url}
\usepackage[english]{babel}
\usepackage[utf8]{inputenc}
\usepackage[normalem]{ulem}
\usepackage{amsmath}
\usepackage{graphicx,epstopdf}
\usepackage{amssymb,epsfig,color,textcase}
\usepackage{hyperref}
\usepackage{doi}
\providecommand{\U}[1]{\protect\rule{.1in}{.1in}}

\begin{document}
\let\WriteBookmarks\relax
\def\floatpagepagefraction{1}
\def\textpagefraction{.001}

\shorttitle{Ground-state structure in double-perovskite PrBaMn$_2$O$_6$}

\shortauthors{SV Streltsov et~al.}


\title [mode = title]{Ground-state structure, orbital ordering and metal-insulator transition in double-perovskite PrBaMn$_2$O$_6$}

\author[1,2]{Sergey V. Streltsov}[orcid=0000-0002-2823-1754]
\cormark[1]
\ead{streltsov@imp.uran.ru}
\author[3]{Roman E. Ryltsev}[orcid=0000-0003-1746-8200]
\author[4]{Nikolay M. Chtchelkatchev}[orcid=0000-0002-7242-1483]
\affiliation[1]{organization={M.N. Miheev Institute of Metal Physics of Ural Branch of Russian Academy of Sciences},
    addressline={S. Kovalevskaya St. 18}, 
    city={Ekaterinburg},
    postcode={620990}, 
    country={Russia}}
 \affiliation[2]{organization={Ural Federal University},
    addressline={Mira St. 19}, 
    city={Ekaterinburg},
    postcode={620002}, 
    country={Russia}}  
 \affiliation[3]{organization={Institute of Metallurgy of the Ural Branch of the Russian Academy of Sciences},
    addressline={Amundsena St. 101}, 
    city={Ekaterinburg},
    postcode={620016}, 
    country={Russia}}
 \affiliation[4]{organization={Vereshchagin Institute for High Pressure Physics, Russian Academy of Sciences},
    addressline={Kaluzhskoye shosse, 14}, 
    city={Troitsk, Moscow},
    postcode={142190}, 
    country={Russia}}
    
\cortext[cor1]{Corresponding author}

\begin{abstract}
In recent years, A-site ordered half-doped double-perovskite manganites $\rm RBaMn_2O_6$ (R=rare earth) have attracted much attention due to their remarkable physical properties and a prospect of application as magnetoresistance, multiferroic, and oxygen storage materials. The nature of the ground state in ${\rm RBaMn_2O_6}$ as well as sequence of phase transitions taking place at cooling are not yet well understood due to complexity in both experimental and theoretical studies. Here we address the origin of the ground-state structure in PrBaMn$_2$O$_6$ as well as its electronic and magnetic properties. Utilizing GGA+U approach and specially designed strategy to perform structural optimization, we show that the system has two competing AFM-A and AFM-CE magnetic structures with very close energies. The AFM-A structure is a metal, while AFM-CE is an insulator and the transition to the insulating state is accompanied by the charge Mn$^{3+}$/Mn$^{4+}$, and orbital $3x^2-r^2$/$3y^2-r^2$ orderings. This orbital ordering results in strong cooperative Jahn-Teller (JT) distortions, which lower the crystal symmetry. Our findings give a key to understanding contradictions in available experimental data on ${\rm PrBaMn_2O_6}$ and opens up the prospects to theoretical refinements of ground-state structures in other ${\rm RBaMn_2O_6}$ compounds.
 \end{abstract}

\begin{highlights}
\item $\rm RBaMn_2O_6$ has two competing AFM-A and AFM-CE magnetic structures
\item AFM-A structure is a metallic while AFM-CE is an insulating
\item Metal-Insulator transition  is accompanied by charge (Mn$^{3+}$/Mn$^{4+}$) and orbital ($3x^2-r^2$/$3y^2-r^2$) orderings
\item Strong cooperative Jahn-Teller distortions in AFM-CE configuration lead to P2$_1$/c structure
\end{highlights}



\maketitle

\section {Introduction \label{Intro}}



Perovskite-type oxides are common inorganic compounds, which have attracted attention for decades due to their remarkable physical properties and a wide range of possible applications\cite{He2020,Nan2019,Dai2020,Pena2001}. These properties can be further improved and tuned by fabricating double perovskite structures with general chemical formula  AA'BB'O$_6$ (where A/A' are alkaline/rare earth metals, B/B' are transition/non-transition metals). Due to differences in atomic radii of  A/A' and/or B/B' ions and possibility of their ordering in alternating perovskite layers, such compounds demonstrate complicated behavior and can serve as a basis for new multifunctional materials~\cite{ElBaggari2021,Sheikh2020,Shaikh2021,Pang2020,Zhu2020,Liu2021,Zheng2022,Muscas2022,Bondzior2021,Tahani2022}.



    Half-doped double-perovskite manganites with the formula ${\rm B^{3+}A^{2+}Mn_2O_6}$ reveal complicated interplay between charge, orbital, and spin degrees of freedom, which is mostly caused by a mixed-valence state of the manganese ions and strong correlations between Mn-3$d$ and O-2$p$ orbitals. Among such compounds, $\rm RBaMn_2O_6$ ones, where R=rare earth, are of special interest. Due to the large difference in atomic radii of Ba and R, these materials demonstrate pronounced magnetoresistive and magnetocaloric effects, and relatively high temperatures of magnetic phase transitions, which makes them promising candidates for practical uitilization \cite{Ueda2007,Yamada2017,Yamada2019,Nowadnick2019,Sagayama2014,Mero2019, Blasco2021,Blasco2021JPhysChemC,Sterkhov2022}. 

\begin{figure*}[t]
\centering
\includegraphics[width=0.8\textwidth]{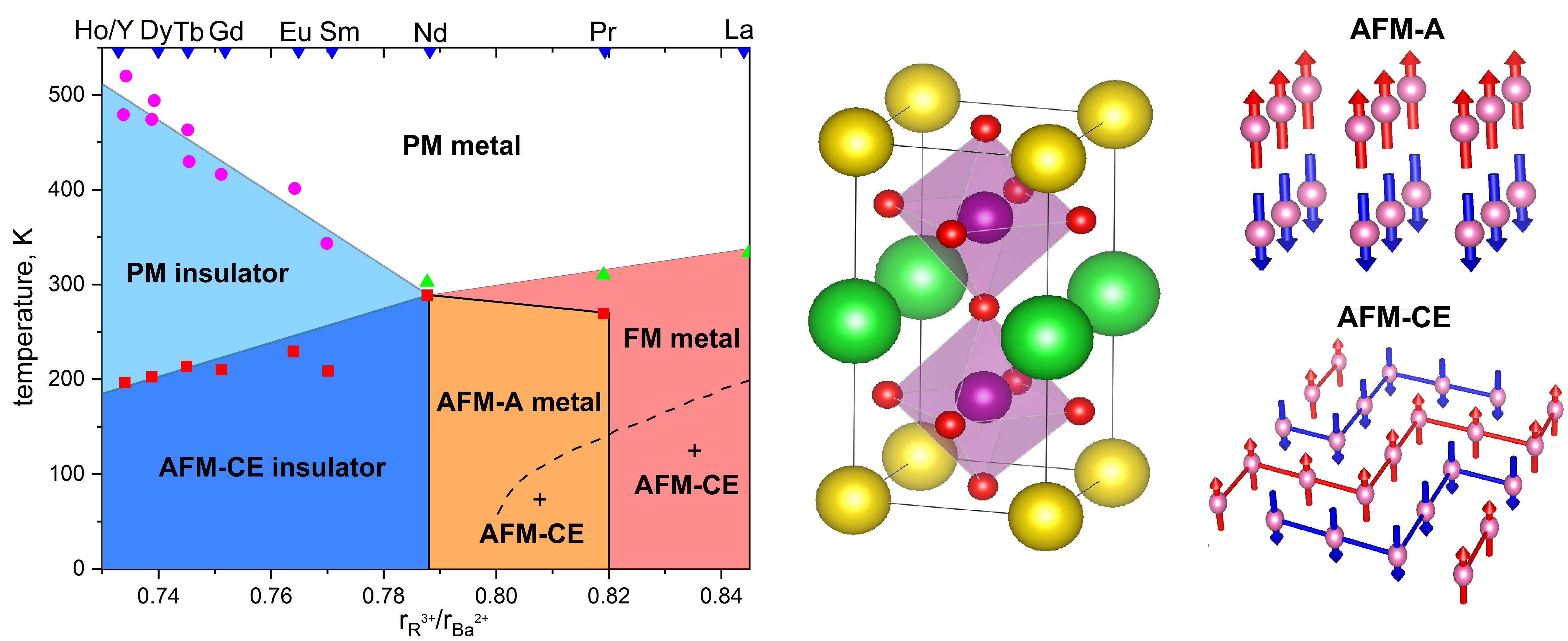}
\caption{\label{fig:PD} Left: Experimental phase diagram of the A-site ordered double-perovskite ${\rm RBaMn_2O_6}$ manganites (R = rare earth), adapted from Refs.~\cite{Nakajima2002JPCS,Nakajima2002,Nakajima2003}. Here CO = charge order, OO = orbital order, PM = paramagnetic, FM = ferromagnetic, AFM = antiferromagnetic. Symbols represent experimental data; solid lines are guide to the eyes; dashed line separates the area where the presence of second AFM-CE phase was detected by neutron scattering experiments. Middle: The unit cell of ${\rm RBaMn_2O_6}$ compounds, where Mn is shown by purple, O by small red, Ba by green, and R by yellow balls. Right: spatial spin distributions in AFM-A and AFM-CE magnetic structures.}
\end{figure*}

Despite an ongoing intensive research, the nature of the ground state in ${\rm RBaMn_2O_6}$ compounds as well as low temperature phase transitions are not yet well understood.  As was suggested in a series of pioneering experimental works by Nakajima et al.~\cite{Nakajima2002JPCS,Nakajima2002,Nakajima2003}, an increase of the rare-earth ionic radius causes the change of the ground state from a CE-type antiferromagnetic (AFM-CE) insulator  to an  to a ferromagnetic (FM) metal through an intermediate A-type AFM (AFM-A) metal, see Fig.~\ref{fig:PD}. However, recent studies  revealed that for R=(Nd, Pr) the situation can be more complicated~\cite{Yamada2017,Yamada2019,Mero2019,Blasco2021,Sterkhov2022}. Investigations of powder $\rm (Nb, Pr)BaMn_2O_6$ samples showed that they transform under cooling from high-temperature paramagnetic phase to ferromagnetic metal and  subsequently to  the AFM-A phase~\cite{Nakajima2002,Nakajima2003}. Moreover, the low-temperature FM-AFM phase transitions are accompanied by a change of lattice parameters –- elongation of the $a$ axis and the contraction of the $c$ axis. It was observed that during all the transitions the systems keep their high-temperature tetragonal P4/mmm symmetry. However, it was later shown~\cite{Yamada2017} that single-crystal $\rm NbBaMn_2O_6$ undergoes a first-order metal-insulator transition at $T_{\rm MI} \approx 290$ K accompanied by the change of the symmetry to $P2_1am$. Besides, the N\'eel temperature of the single crystals is about 50 K lower than for the powder samples. In the case of $\rm PrBaMn_2O_6$ the situation is even more intricate. Initially, Nakajima et al. suggested that structural phase transition is coupled with FM-AFM transition~\cite{Nakajima2002,Nakajima2003}. However, as was shown later these transitions occur at different temperatures~\cite{Titova2020}. Recently some of us studied the origin of structural transition in  $\rm PrBaMn_2O_6$ and revealed that the transition is caused by  the splitting of $e_g$-doublet due to the tendency towards orbital ordering at the metal-insulator transition~\cite{Sterkhov2022}. However, no conclusive evidences of this mechanism  have been obtained. Results for X-ray powder diffraction confirmed the conclusions of Nakajima et al. that high-temperature P4/mmm symmetry survives down to at least 150 K. This contradicts recent results by Blasco et al.~\cite{Blasco2021} who performed a structural study of $\rm PrBaMn_2O_6$ using synchrotron radiation X-ray powder diffraction and Raman spectroscopy and argued that the structural transition is accompanied with the change of crystal symmetry to $P2_1am$. Moreover, this transition is coupled to an electronic localization and an antiferromagnetic transition~\cite{Blasco2021}.

The key to understanding the above contradictions in $\rm PrBaMn_2O_6$  probably lies in the results of neutron diffraction experiments~\cite{Nakajima2003}, which revealed that, at low temperatures, besides AFM-A phase, some portion of AFM-CE phase is observed (see dashed line in Fig.~\ref{fig:PD}). This suggests that $\rm PrBaMn_2O_6$ has two competing magnetic phases whose energies are very close. Therefore, at low temperatures, $\rm PrBaMn_2O_6$ samples   are probably composed of a mixture of these phases. Unfortunately,  both atomic and electronic structures of these phases are not clear. Particularly, it is not clear  if these phases are responsible for metal-insulator transition. It is rather difficult to  address these issues experimentally.  Indeed, both X-ray and neutron diffraction experiments performed on powder samples can hardly separate contributions of two phases with similar structures. Single-crystals experiments might be more reliable but the fabrication of high-quality samples is an extremely difficult task.

In this situation, reliable {\it ab initio} calculations may help to shed light on the ground-state structure of the systems under consideration as well as to their electronic and magnetic properties. In this paper we utilize GGA+U approach and specially designed strategy to perform structural optimization,  see Sec.~\ref{Strategy}, to uncover the origin of the low-temperature insulating state in PrBaMn$_2$O$_6$ compound as well as its magnetic and crystal structure in the ground state. 

\section {Calculation details}

The calculations were performed within the density functional theory (DFT) using the generalized gradient approximation (GGA)~\cite{Perdew1996} and taking into account strong Coulomb interactions via DFT+U~\cite{Anisimov1997} scheme with Hund's intra-atomic exchange $J_H=1$ eV and Hubbard $U=6$ eV for Mn, as was estimated by constraint DFT~\cite{Aryasetiawan2006}. We utilized VASP for all calculations~\cite{Kresse1996}. Pr-$4f$ states were treated as a core. The calculations were carried in a large supercell  consisting of 80 atoms to allow for different types of magnetic order. The Monkhorst-Pack scheme~\cite{Monkhorst1976} was used for integration of the Brillouin zone with 32 $k-$points in the supercell. The symmetry constraints  were not taken into account to allow any possible type of orbital structure to develop.

\section {Strategy \label{Strategy}}

 Jahn-Teller systems are very different from conventional transition metal oxides from the point of view of crystal structure optimization. In these systems there can be many extremums of energy surface separated from each other by  substantial energy barriers  even for an isolated Jahn-Teller ion. The situation gets more complicated, when we have a solid consisting of JT ions and need to take into account not only  dominant factors such as correlation effects, spin-orbit coupling and the exchange interaction, but also vibronic coupling with all possible phonon modes. In such a case, a straightforward DFT relaxation often fails to find the correct ground state and gets stuck in the local minima. 

This  exact situation is realized in PrBaMn$_2$O$_6$ if one begins with the low-temperature crystal structure obtained from available X-ray powder diffraction data \cite{Sterkhov2021}. The resulting  relaxed structure neither  exhibits large difference in transition metal-ligand bond lengths expected for the Jahn-Teller active Mn ion, nor leads to the  observed insulating ground state.
\begin{figure}[t!]
\includegraphics[width=0.49\textwidth]{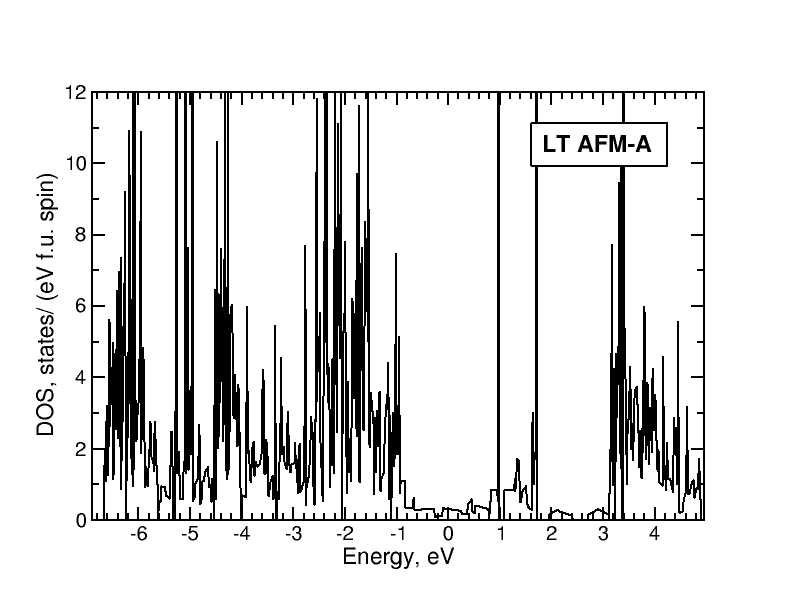}\\
\includegraphics[width=0.49\textwidth]{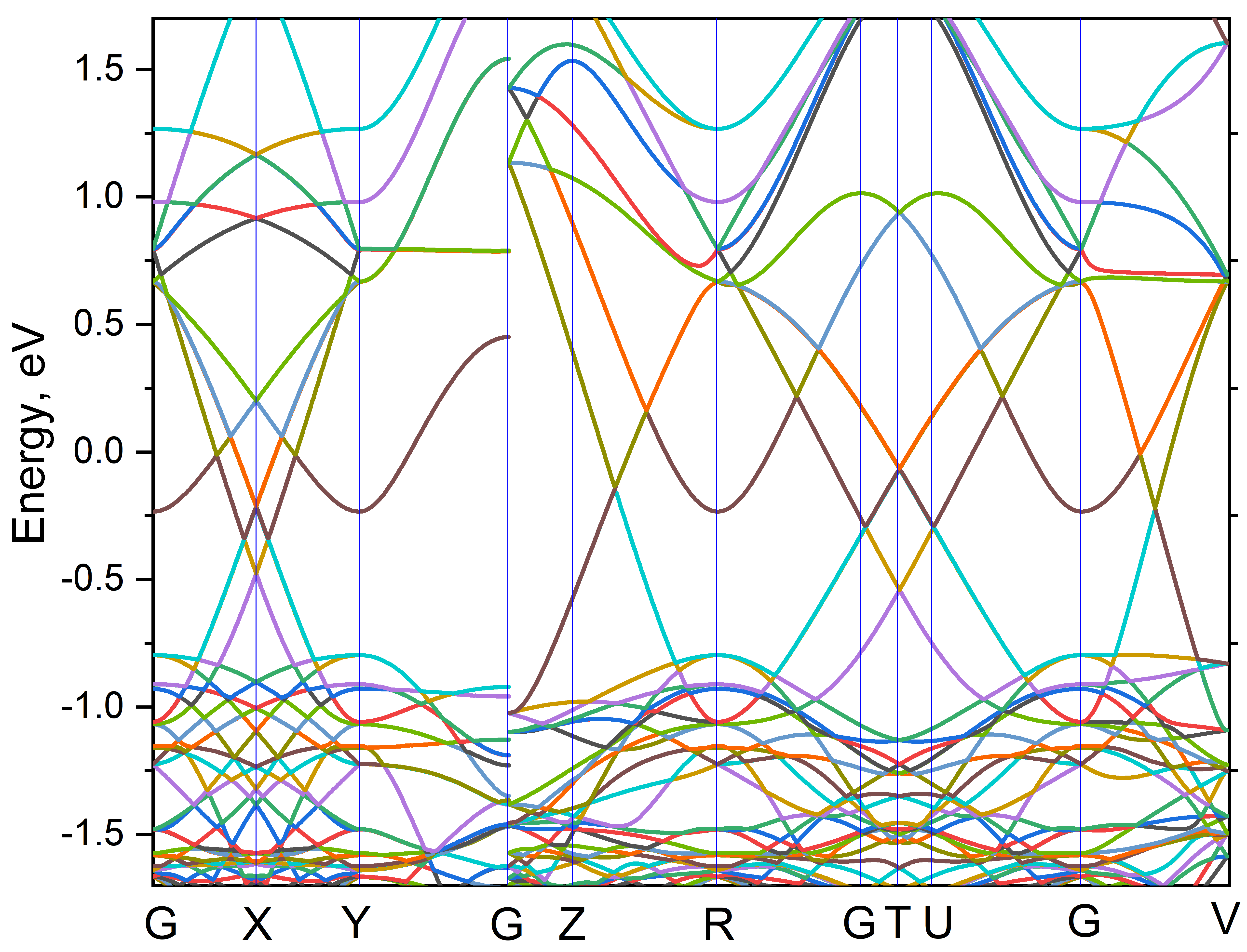}
\caption{\label{DOS-A} Total density of states and band structure for AFM-A$_{ab}$ magnetic structure calculated for lattice constants corresponding to low-temperature (185 K) phase. Fermi energy is in zero.}
\end{figure}

The recommended strategy in this case is to take the symmetrized available low-temperature or high-temperature  crystal structure, (changing the volume accordingly). Then one either assumes a reasonable magnetic structure~\cite{Streltsov2014d} or one  needs to slightly shift  the ligands according to expected Jahn-Teller distortions \cite{Streltsov12Cu}. 

In case of PrBaMn$_2$O$_6$ we used the first strategy. The microscopic mechanism lying beyond this procedure is a strong interplay between spin and orbital degrees of freedom - Kugel-Khomskii coupling. In DFT+U typically various magnetic configurations can be easily stabilized. Then the superexchange interaction \cite{Goodenough,Khomskii2020} results in a particular orbital ordering, and  ultimately the lattice  responds to this orbital pattern developing corresponding distortions. Thus, one can begin with a particular magnetic structure setting up initial magnetic moments according to the charge state of each Mn, i.e. 3 or 4 $\mu_B$.

\section {Results \label{DFT}}

\subsection{High-temperature structure}
\begin{figure}[t!]
\includegraphics[width=0.49\textwidth]{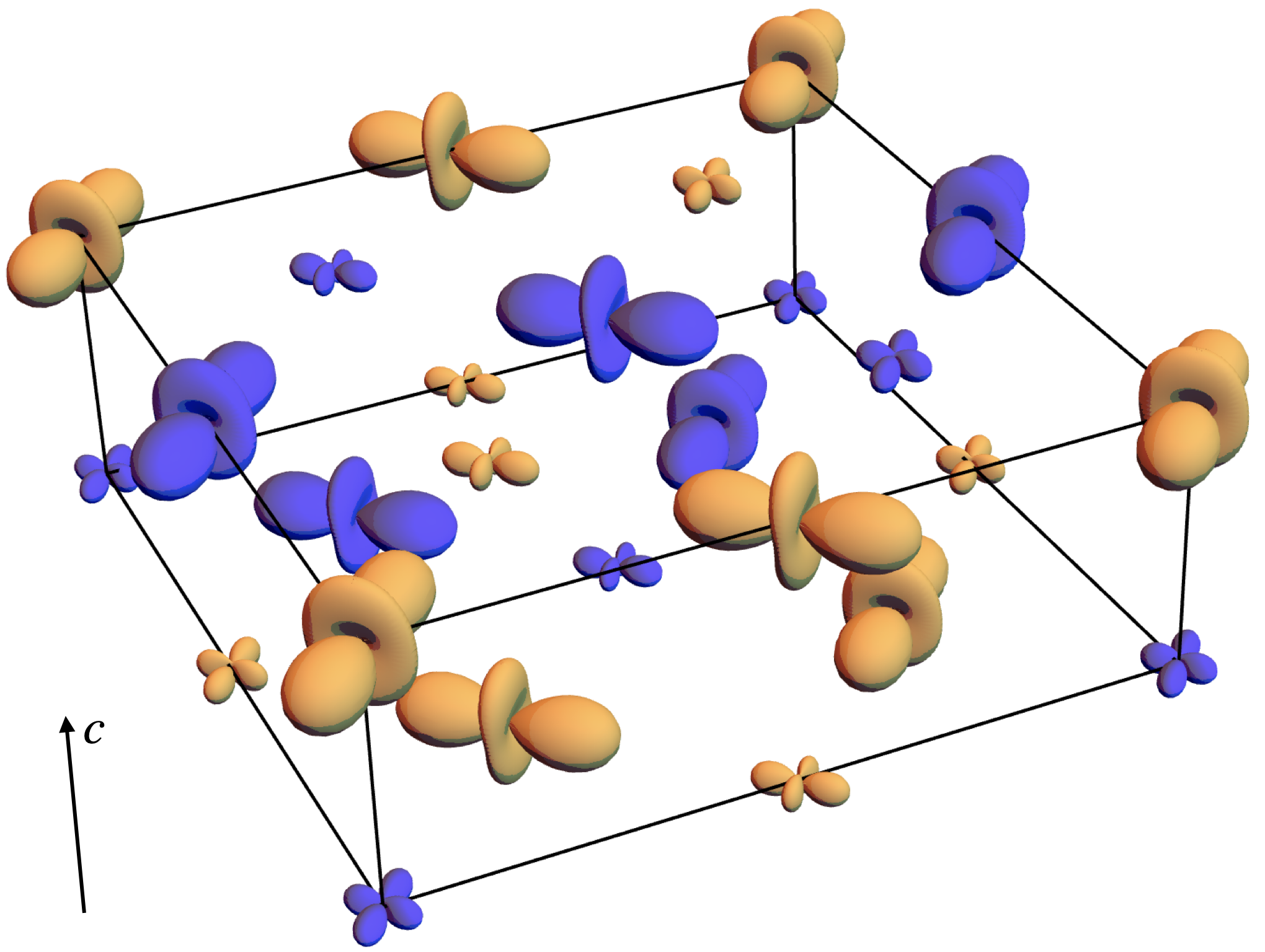}
\caption{\label{OO} The orbital ordering obtained in GGA+U approach with AFM-CE$_{ab}$-OO$_c$ magnetic structure calculated for lattice constants corresponding to low-temperature (185 K) phase. The $e_g$ orbital having the largest occupancy (and normalized by this occupancy) is plotted. One can see that there are two types of Mn ions: Mn$^{3+}$ ($d^4$) having one electron at the $3x^2-r^2$ or $3y^2-r^2$ orbital and  Mn$^{4+}$ ($d^3$) ions, which have only a small admixture at $e_g$ subshell due to hybridization with ligands. Colors correspond to different spin projections.}
\end{figure}

For the sake of completeness we start with experimental high-temperature structure (space group P4/mmm \cite{Mostovshchikova2021}), but increase the unit cell  to 16 Mn ions in order to be able to simulatethe experimentally observed  AFM-CE magnetic structure. This is one of the typical magnetic orderings in manganates, it can be visualized as ferromagnetic zigzag chains in the $ab$ plane with Mn$^{3+}$ ($d^4$) ions at the legs and Mn$^{4+}$ ($d^3$) ions at the corners of these chains, see Fig.~\ref{OO}.  This structure can have FM or AFM ordering along the $c$ direction, these two configurations are denoted as AFM-CE$_{ab}$-FM$_{c}$ and AFM-CE$_{ab}$-AFM$_{c}$ correspondingly. In our consideration we also include pure FM, AFM-A  state, which contains ferromagnetic planes coupled antiferromagnetically and  the so-called AFM-CE$_{ab}$-OO$_{c}$ configuration. 
\begin{table}[b]
\centering
\caption{Total energies (in meV per formula unit) of different magnetic structures for the high-temperature (HT) structure. The crystal structure was kept undistorted (second column) or allowed to be relaxed (third column).}
\begin{tabular}{l c c c}
\hline
\hline
 Magnetic structure & Total energy, & Total energy,\\
    & no relaxation &  relaxed\\
\hline 
  FM                      & 0    & 0\\
  AFM-A$_{ab}$            & 22.7  & 22.7\\
  AFM-CE$_{ab}$-FM$_{c}$ & 181.9 & 45.8\\
  AFM-CE$_{ab}$-AFM$_{c}$ & 181.8 & 44.4\\
  AFM-CE$_{ab}$-OO$_{c}$  & 174.4 & 10.3 \\
\hline
\hline
\end{tabular}
\label{HT}
\end{table}

The  AFM-CE$_{ab}$-OO$_{c}$ configuration has Mn$^{4+}$ ions on top and bottom of Mn$^{3+}$ ions along the $c$ axis. In this case  zigzags in the different $ab$ planes get shifted. The idea to include this configuration is related to the fact that, if we have Mn$^{3+}$ with an electron on e.g. the $3x^2-r^2$ orbital, then ligands in the perpendicular plane, i.e $yz$, tend to compress this octahedron. In  the AFM-CE$_{ab}$-FM$_{c}$ and AFM-CE$_{ab}$-AFM$_{c}$ configurations two Mn$^{3+}$ ions in the different $ab$ planes on top of each other have a common oxygen that can not participate in compression of both octahedra and thus we are unable to fully minimize elastic contributions. In contrast, in  the AFM-CE$_{ab}$-OO$_{c}$ solution ligands are free to follow the orbital ordering not only in the $ab$ plane, but also along the $c$ direction. This  is important for  DFT calculations where atomic relaxation is allowed.

Calculated  total energies of these magnetic structures in case of the HT structure are summarized in Tab.~\ref{HT}. One can see that the ground state corresponds to the FM configuration. This agrees with experimental findings:  upon cooling the  paramagnetic  state orders exactly to FM structure ~\cite{Nakajima2003,Nakajima2004}, see also Fig.~\ref{fig:PD}. 

 The results for a case when the lattice is allowed to change are shown in the last column in Tab.~\ref{HT}.  In this calculation the ground state stays ferromagnetic but the  the state closest in energy turns out to be  the AFM-CE$_{ab}$-OO$_{c}$ state, which was designed to gain energy due to the Jahn-Teller distortions of Mn$^{3+}$O$_6$ octahedra both in the $ab$ plane and along the $c$ direction. Analysis of the occupation numbers demonstrates that in this case we indeed have the charge ordering, such that Mn ions at the legs of zigzags are 3+ (four large eigenvalues of the occupation matrix close to 0.9) and those in the corners
are 4+ (three eigenvalues $\sim$ 0.9). Moreover, there is 
also the ordering of $3x^2-r^2$ and $3y^2-r^2$  orbitals on Mn$^{3+}$ sites. This results in a strong ferromagnetic exchange due to overlap between half-filled and empty $e_g$ orbitals~\cite{Streltsov-UFN}, which supports CE type of magnetic structure. The coupling between chains is antiferromagnetic because of the half-filled $t_{2g}$ orbitals.  Notably,  the onset of the charge and orbital order is enough for stabilization of the insulating state with the band gap  around 0.9 eV. Both FM and AFM-A$_{ab}$ solutions turn out to be metallic without any charge ordering. 

\subsection{Low-temperature structure}
Next, calculations  of the same type were performed to simulate  the low-temperature (LT) structure. We, again, started with the undistorted HT structure, but with the $a$ and $c$ parameters changed to reproduce the experimental unit cell at 185~K, i.e. $a$ was increased by 0.6\%, while $c$ was reduced by 1.2 \% \cite{Sterkhov2021}.

Remarkably, this turns out to be enough to switch  a sign of the exchange interaction between the $ab$ planes and change the magnetic ground state from FM to AFM-A$_{ab}$. Most probably this is due to  the enhanced antiferromagnetic superexchange interaction along the $c$ direction. The system retains P4/mmm space group with AFM-A$_{ab}$ magnetic structure and does not  exhibit neither charge nor orbital ordering with all Mn having magnetic moment 3.55$\mu_B$ and the metallic ground state, see Fig.~\ref{DOS-A}. We also do not observe any indirect band gaps in the band structure of AFM-A$_{ab}$.

\begin{table}[b!]
\centering
\caption{Total energies (in meV per formula unit) of different magnetic structures for the low-temperature (LT) structure. The crystal structure was  allowed to be relaxed.}
\begin{tabular}{l c c c}
\hline
\hline
 Magnetic structure &  Total energy,\\
&    relaxed\\
\hline 
  FM                      &  9.3\\
  AFM-A$_{ab}$            &  0\\
  AFM-CE$_{ab}$-FM$_{c}$ &  41.5\\
  AFM-CE$_{ab}$-AFM$_{c}$ &  29.9\\
  AFM-CE$_{ab}$-OO$_{c}$  &  2.1\\
\hline
\hline
\end{tabular}
\label{LT}
\end{table}

It should be noted that the difference between AFM-A$_{ab}$ and the next in energy AFM-CE$_{ab}$-OO$_{c}$ structure is tiny, $\sim$ 24 K/f.u. Thus, this is not surprising that both phases are experimentally seen in the real material~\cite{Nakajima2003}. Moreover, subtle modifications of the sample's quality, such as exact stoichiometry, degree of disorder, presence or absence of defects in different grains may affect the magnetic ordering and  the crystal structure  strongly coupled to it.  The significance of some of these factors for magnetic properties of RBaMn$_2$O$_6$, where R is a rare-earth ion or Y, has been already mentioned \cite{Nakajima2004}.

One can argue that  the AFM-A structure is stabilized by the double-exchange-like mechanism, which is favoured by wide metallic bands. In contrast, the charge ordering and  the insulating behaviour  associated to it  are important for the AFM-CE state.  The presence of  charge ordering results in stronger local distortions, which leads to a loss in elastic energy but it is outweighed by the gain of energy due to the exchange interaction and local crystal fields.

 The crystal structure  optimized in AFM-CE$_{ab}$-OO$_{c}$ calculations was refined using Findsym~\cite{Stokes2005}. If  a tolerance factor for lattice parameters and for atomic positions are  chosen to be 0.0001 \AA~ and 0.005 \AA~then the resulting structure is  a triclinic $P1$ (atomic positions are given in SI). However, if they are increased to  0.01 \AA~and 0.03 \AA~the structure can be described by  a P2$_1$/c space group with atomic positions presented in Tab.~\ref{CE-crystal-structure} (note that these calculations take into account chosen type of magnetic structure). It has to be mentioned that this result strongly differs from P2$_1am$ structure previously proposed  for NdBaMn$_2$O$_6$~\cite{Yamada2017} and PrBaMn$_2$O$_6$~\cite{Blasco2021}. In fact, there is only one structural type of Mn ions in P2$_1am$ and, therefore it is inconsistent with AFM-CE demanding two different positions for Mn$^{3+}$ and Mn$^{4+}$  ions.
 \begin{figure}[t]
\includegraphics[width=0.49\textwidth]{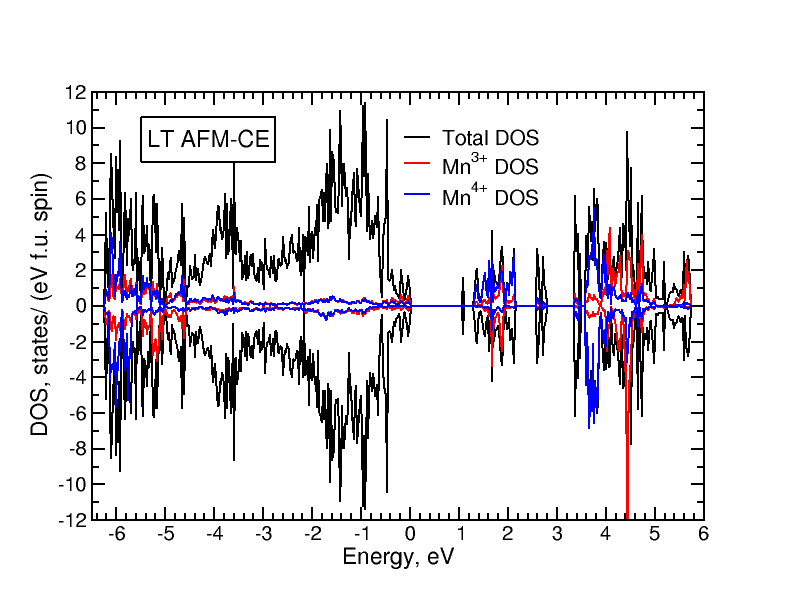}
\caption{\label{DOS} Total and partial density of states for AFM-CE$_{ab}$-OO$_{c}$ magnetic structure calculated for lattice constants corresponding to low-temperature (185 K) phase. Fermi energy is in zero.}
\end{figure}

These two types of Mn sites are clearly seen in both P1 and P2$_1$/c structures. The first type of sites has strongly elongated ligand octahedra with one long ($\sim 2.09$ \AA) and four short ($\sim 1.93$ \AA) Mn-O bonds. This site is occupied by Jahn-Teller active Mn$^{3+}$. The second one is not stretched along one of the directions, but all Mn-O bonds are still different and vary from 1.89 to 2.02 \AA~ since they must comply  to distortions induced by Mn$^{3+}$ ions. This site accommodates Mn$^{4+}$ ions. Since both sites have $C_1$ point group (being refined in P2$_1$/c structure), which does not include inversion center, one might expect a ferroelectric response, which should  be seen e.g. in optical measurements. This behaviour was found in quadruple Mn perovskites such as e.g. 
BiMn$_7$O$_{12}$~\cite{Imamura2008,Mezzadri2009} and CaMn$_7$O$_{12}$~\cite{Zhang2011,Johnson2012}.

Finally, in Fig.~\ref{DOS} we present the electronic structure of PrBaMn$_2$O$_6$ obtained in  the AFM-CE$_{ab}$-OO$_{c}$ configuration. The band gap is $\sim$ 1 eV. One should keep in mind that it depends on the choice of Hubbard $U$, but this value agrees with optical measurements by Mostovshikova et al. demonstrating that the band gap is larger than 0.55 eV~\cite{Mostovshchikova2021}. Top of the valence band is formed by O-$2p$ states, while bottom of the conduction band is composed of Mn-$e_g$ orbitals. Magnetic moments were found to be partially suppressed due to strong hybridization with ligands: 3.7$\mu_B$  for Mn$^{3+}$ and 3.2 $\mu_B$ for Mn$^{4+}$, which is typical for manganates (see e.g. \cite{Streltsov2014d}). 

Thus, the metal-insulator transition observed experimentally in PrBaMn$_2$O$_6$  can be related to the formation of  the insulating AFM-CE phase. This insulating phase is clearly observed in RBaMn$_2$O$_6$ systems with small rare-earth ions, see Fig.~\ref{fig:PD}.  For larger rare-earths (NdBaMn$_2$O$_6$, PrBaMn$_2$O$_6$, and LaBaMn$_2$O$_6$) the situation is somewhat more complicated due to presence of additional AFM-A phase. However, one  should also expect that conductivity in these materials will be strongly suppressed because of  larger disorder, correlation effects and  the presence of the insulating AFM-CE phase.

\begin{table}[b!]
\centering
\caption{Optimized crystal structure corresponding to AFM-CE$_{ab}$-OO$_{c}$ configuration refined with tolerance factor 0.01 \AA~for the lattice parameters and 0.03 \AA~for atomic positions. Space group was found to be P2$_1$/c with lattice parameters $a=7.65530$ \AA, $a=5.55305$ \AA,~and $a=11.10610$ \AA~ and $\beta=90^{\circ}$.}
\begin{tabular}{l c c c}
\hline
\hline
 Ion &  & Position & \\
\hline 
Mn1 & 0.24988 & 0.75453  &0.37487 \\
Mn2 & 0.24384 & 0.24240  &0.62461 \\
O1  & 0.00084 & 0.78900  &0.39644 \\
O2  & 0.25514 & -0.00884 &0.50035 \\
O3  & 0.23147 & 0.01624  &0.00477 \\
O4  & 0.21107 & 0.47843  &0.74355 \\
O5  & 0.21377 & 0.51343  &0.25269 \\
O6  & 0.50403 & 0.74877  &0.36438 \\
Pr  & -0.00074 & 0.24965 & 0.37856 \\
Ba  &  0.49923 & 0.25536 & 0.37613 \\
\hline
\hline
\end{tabular}
\label{CE-crystal-structure}
\end{table}

\section {Conclusions}
In the present paper, the origin of the insulating state at low temperatures, magnetic state and crystal structure of one of double manganates PrBaMn$_2$O$_6$ were studied. We show that  the AFM-A and AFM-CE magnetic structures are very close in energy and such factors as stoichometry, defects of the crystal structure or disorder may favor one of these orders. This is consistent with neutron powder diffraction data detecting both structures in the low-temperature phase of PrBaMn$_2$O$_6$~\cite{Nakajima2003}. 

 The AFM-A structure was found to be metallic  in our calculations, while the AFM-CE  state is  an insulator and the transition to the insulating state occurs hand in hand with the charge (Mn$^{3+}$/Mn$^{4+}$) and orbital ($3x^2-r^2$/$3y^2-r^2$) orderings. This orbital ordering results in strong cooperative Jahn-Teller distortions, which lower the crystal structure symmetry.  The GGA+U calculations allowed us to predict the crystal structure for the low-temperature phase of PrBaMn$_2$O$_6$. This result can be also relevant for many other manganates RBaMn$_2$O$_6$  which have the double perovskite structure and the AFM-CE type of magnetic  order. We  believe that it will also assist the refinement of available and future X-ray and neutron diffraction data.

\section {Acknowledgments}
We are extremely grateful to P. Maksimov for his comments and to E. Sterkhov providing us with results  of high-temperature refinement of crystal structure.

We acknowledge support of the Russian Science Foundation via projects 20-62-46047 (study of interplay between charge, spin and orbital degrees of freedom) and 18-12-00438 (structure optimization). Data processing was supported by RFBR (project No 19-29-12013) and the Quantum project (AAAA-A18-118020190095-4). 

\bibliographystyle{elsarticle-num}
\bibliography{library}

\end{document}